\documentclass{article}
\usepackage{geometry}
\usepackage{natbib}

\usepackage{graphicx}                    
\usepackage{url}  
\usepackage{setspace}
\usepackage{hyperref}

\title{First Joint Observations of Space Weather Events over Mexico}

\author{V. De la Luz$^{1,2}$
  \and
  J.A Gonz\'alez-Esparza$^{2}$
  \and
  M.A. Sergeeva$^{1,2}$
  \and
  P. Corona-Romero$^{1,2}$
  \and
  L.X. Gonz\'alez$^{1,3}$
  \and
  J. Mejia-Ambriz$^{1,2}$
  \and
  J.F. Vald\'es-Galicia$^{3}$
  \and
  E. Aguilar-Rodriguez$^{2}$
  \and
  M. Rodriguez-Martinez$^{4}$
  \and
  E. Romero-Hernandez$^{5}$
  \and
  E. Andrade$^{2}$
  \and
  P. Villanueva$^{2}$
  \and
  E. Huipe-Domratcheva$^{4}$
  \and
  G. Cifuentes$^{2}$
  \and
  E. Hernandez$^{3}$
  \and
  C. Monstein$^{6}$
  \and
  $^{1}$Conacyt - Servicio de Clima Espacial Mexico, Laboratorio Nacional de Clima Espacial, LANCE, Morelia, Mexico.\\
  \and
  $^{2}$Instituto de Geofisica, Unidad Michoacan, UNAM, C.P.58089, Mexico.\\
  \and
  $^{3}$Instituto de Geof\'isica, Universidad Nacional Aut\'onoma de M\'exico, CDMX, M\'exico.\\
  \and
  $^{4}$Escuela Nacional de Estudios Superiores unidad Morelia, UNAM, Michoac\'an, M\'exico.\\
  \and
  $^{5}$Universidad Autónoma de Nuevo Le\'on, FCFM, LANCE, Mexico.\\
  \and
  $^{6}$ Institute for Astronomy, Institute for Particle Physics and Astrophysics, Zürich, Switzerland.
}

\begin{document}
\maketitle

\begin{abstract}
  The Mexican Space Weather Service (SCiESMEX in Spanish) and National Space Weather Laboratory (LANCE in Spanish) were organized in 2014 and in 2016 respectively
  to provide
  space weather monitoring and alerts, as well as scientific research in Mexico. In this work, we present the results of the first joint observations of two events (22 June, 2015, and 29 September, 2015) with our local network of instruments and their related products. This network includes the MEXART radio telescope (solar flare and radio burst), the Compact Astronomical Low-frequency, Low-cost Instrument for Spectroscopy in Transportable Observatories (CALLISTO) at MEXART station (solar radio burst), the Mexico City Cosmic Ray Observatory (cosmics ray fluxes), GPS receiver networks (ionospheric disturbances), and the Geomagnetic Observatory of Teoloyucan (geomagnetic field). The observations show that we detected significant space weather effects over the Mexican territory: geomagnetic and ionospheric disturbances (22 June, 2015), variations in cosmic rays fluxes, and also radio communication's interferences (29 September, 2015). The effects of these perturbations were registered, for the first time, using space weather products by SCiESMEX: TEC maps, regional geomagnetic index K$_{\rm mex}$, radio spectrographs of low frequency, and cosmic rays fluxes. These results prove the importance of monitoring space weather phenomena in the region and the need to strengthening the instrumentation network.
\end{abstract}


  \section{Introduction}
Space weather (SW) phenomena influence the performance and reliability of different modern technological systems; see for instance \cite{1999SSRv...88..563B} and \cite{Dinardini2016a}.
  The country has developed a significant infrastructure that is vulnerable to SW events, such as electricity generation and a transportation grid, telecommunications, electronic banking, long pipelines for gas and oil transportation, etc. The effects of the Carrington geomagnetic storm in 1859 were registered in several locations, indicating that the region is vulnerable to extreme geomagnetic storms  \citep{gonzalez2018}.

There are some studies of particular SW events that affected the geomagnetic field and ionosphere in Mexico, for example \cite{Rodriguez-Martinez2014}, \cite{2015AdSpR..55..586L}, \cite{SERGEEVA2017}, \cite{ROMEROHERNANDEZ2017}, and \cite{SERGEEVA2018}); however,
the SW phenomena in this region have not been studied comprehensively. For instance, there is a lack of continuous multi-instrument observations of SW phenomena in Mexico that can provide reliable statistics for regional SW studies. 
Mexico is situated at low latitudes (geographic latitudes 14$^{\circ}$- 32$^{\circ}$N, geomagnetic latitudes 23$^{\circ}$ - 38$^{\circ}$ N).
Recent studies
prove that the SW effects are far from
being fully understood
at these latitudes \citep{2014JSWSC...4A..28C, SWE:SWE20063, balch2004halloween,balch2004intense}. 

The 
southern half
of Mexican territory is located between the Northern Tropic and the Equator. The Sun's incident ray path, at maximum elevation, remains troughout the year between 35$^{\circ}$ and 81$^{\circ}$ in the northern region of the country (Tijuana at 32$^{\circ}$N) and between 53$^{\circ}$ and 90$^{\circ}$ in the southern region
(Tapachula at 14$^{\circ}$N). These conditions 
match countries with 
similar
latitudes such as north of
Africa, the Arabian Peninsula and south of Asia (including south of China and India).  The Sun's paths for these latitudes increase the exposition time of the solar projection over the ground. Consequently, this raises the probability of radio interferences detected at ground level, produced directly or indirectly by a solar radio burst, Solar Energetical Particles (SEPs), and flares 
\citep{Lanzerotti2007}. Models like D-Region Absorption Predictions (D-RAP) by NOAA show this effect \citep{swpcRAP}.

  Since 2014, Mexico as begun a strategy for SW awareness. In 2014, the
  Mexican Space Weather Service (SCiESMEX) was created;
  in 2016, the National Space Weather Laboratory (LANCE) and the 
  Repository of Space Weather Data (RICE) were established \citep{SWE:SWE20412}.
  Some of the ground-based instruments involved in the SW observations in Mexico have been used for more than 50 years \citep{Dinardini2016a,Dinardini2016b,Dinardini2016c}. Currently, the instrumental network provides
  the possibility of measuring local geomagnetic
field variations, cosmic ray flux, solar wind parameters using interplanetary scintillation
data, solar radio bursts, radio interferences, GPS signals delays, etc.

  
The aim of this work is to estimate
the impact of SW phenomena
over 
Mexico. We based our results on multilateral observations performed by the SCiESMEX instrumental network.
In this work, we addressed two events registered over Mexico by SCiESMEX in 2015: on 22 June, and on 25-29 September. The first event was mainly related to a M6.5 solar flare and a geomagnetic storm that caused ionospheric perturbations. The second event was related to a solar radio burst. The paper is organized as follows: Section \ref{instrumentation} introduces the network of SW instruments in Mexico. Section 3 discusses the results of observations during the two events, and final remarks are given in the Conclusions section.

\section{Space Weather Instrumental Network and Products}\label{instrumentation}

This section 
describes the ground-based 
facilities for SW observations 
in Mexico.



\subsection{The Mexican Array Radio Telescope MEXART} 
The Mexican Array Radio Telescope MEXART 
is a transit instrument 
dedicated to 
interplanetary scintillation (IPS) observations from compact radio sources \citep{gonzalez2004}. The instrument has 16 fixed latitudinal beams pointing towards different declinations and uses the Earth's rotation to scan the whole sky. The angular width of the beams along the east-west direction is about one degree, so a discrete radio source has a transit of about four minutes in the data series and the Sun around 8 minutes. The basic elements of the radio telescope are full wavelength dipoles ($\lambda =2.14$m). The radiotelescope performs observations at the frequency of 139.65 MHz with the bandwidth of 2 MHz. More details of the instrument can be found in \cite{2010SoPh..265..309M}.

The telescope allows us to 
remotely infer some characteristics of solar wind streams crossing along the line of sight of the extragalactic radio sources detected by the instrument. This includes the tracking of large-scale interplanetary perturbations. The solar wind speeds and interplanetary density fluctuations along the lines of sights are computed with use of the methods developed by SCiESMEX. When the line of sight of a radio source passes across the solar wind electronic density inhomogeneities, the radio signals are scattered, and a diffraction pattern is produced. 

To infer some solar wind characteristics (velocity and density fluctuations) from the IPS data, we apply a power spectra analysis to record the transit of the radio source. We employ a theoretical model to obtain a power spectrum of the IPS fluctuations. This IPS theoretical spectrum incorporates different physical parameters, including solar wind speed. We fit the theoretical model to the observed power spectra, obtaining the solar wind speed that best matches the observation. The solar wind speed location is assumed at the nearest point of the Sun to the line of sight. The solar wind density fluctuations are estimated from the area under the curve of the observed IPS spectrum; this area is equivalent to scintillation index m. Further details  about the MEXART methodology can be found in \cite{2015SoPh..290.2539M}. The results of IPS observations are published weekly in the SW reports \footnote{Reports can be found in the official webpage of SCiESMEX (\url{http://www.sciesmex.unam.mx/blog/category/reporte-semanal-de-clima-espacial/}).}.




\subsection{CALLISTO Station at MEXART Site}

Solar radio bursts are spontaneous emissions
of electromagnetic waves at low frequencies in the outer solar atmosphere produced by shock waves close to the corona or in the interplanetary medium \citep{2008SoPh..253....3N}. The signal received on Earth is interpreted as an increase of the radio noise
that is
analyzed continuosly by many solar radio spectrum observatories \citep{2012SoPh..277..447I}. The global e-CALLISTO network is among these observatories \citep{2005SoPh..226..143B}.

In 2015, a station of 
CALLISTO was installed
in the facilities of the MEXART radio telescope (CALLISTO-MEXART station). This station forms part of the 
e-CALLISTO network.
Up to now, about 100
solar radio events 
have been detected,
and their radio noise spectrum at the site has been categorized \citep{huipe2018}.
The observations are performed only during the day-light hours with  200 different channels ranging from 45.7 MHz to 344.7 MHz captured every 15 minutes with the resolution of $250$ ms.


The product related to CALLISTO is the dynamic radio spectrograph; see example in Section \ref{mexcal}. Its results are published every 15 minutes at the webpages of RICE and e-CALLISTO international network.

\subsection{Cosmic Ray Observatory}
The Mexico City Cosmic Ray Observatory is equipped with two instruments: a muon telescope and a neutron
monitor (NM). The muon telescope detects the hard component (negative and positive muons) produced
by the
decay of
charged pions,
which are produced by
interactions of the primary cosmic rays with the atmospheric
nucleus.
It is 
composed of eight plates of plastic scintillator; 
four plates are located above the NM and four under it. The muons crossing through
the scintillators lose energy by ionization and produce fluorescent radiation that travels to a
photomultiplier \citep{2010JASTP..72...38A,2007PAN....70.1088S}. The Mexico City NM type is a 6-NM64. The monitor has worked continuously since 1990.
According to \cite{2000SSRv...93..335C}, the mean energy response for the Mexico City NM is 24.5~GV; thus the instrument detects the low energy component of the galactic cosmic rays.

The cosmic ray observatory can detect flux variations caused by solar activity, for example, 
when a coronal mass ejection (CME) strikes the Earth and produces a sudden reduction in galactic cosmic ray flux intensity. This kind of event is known as a Forbush decrease (FD) \citep{1938TeMAE..43..203F}. The cosmic ray intensity may have a drastic decrement up to 20\% in a few hours and the recovery is slow, typically around seven to 10 days. It is one of the extreme manifestations of a transient modulation of Galactic Cosmic Rays (GCRs). FDs are generally correlat with stream interaction regions, interplanetary shocks and/or CMEs originated from the Sun \citep{1996JGR...10121561C,2000SSRv...93...55C,2011SoPh..270..609R}. The FDs are observed by ground-based particle detectors, such as an NM. The phenomenon is produced by the irregularities in the interplanetary magnetic field associated with these large-scale solar wind disturbances that deflect the cosmic ray flux, causing a reduction in the amount of cosmic rays detected at ground level \citep{1971SSRv...12..658L,2015ApJ...814..136G}. The product developed by SCiESMEX computes the cosmic rays flux in real time. The results are included in the SW weekly report.


\subsection{Geomagnetic Observatory}
The Geomagnetic Observatory of Teoloyucan (TEO) is located near Mexico City (at latitude 19.746 N and longitude 99.19 W) and is managed by the Magnetic Service at the Geophysics Institute, UNAM. It performs the measurements of the geomagnetic field components ($D$, $H$, and $Z$) with a local magnetometer.
The sampling rate is five seconds integrated by one minute, and the baseline is continuously calibrated by the Magnetic Service's work team.


Since 2017, SCiESMEX, in collaboration with the Magnetic Service, has estimated the local geomagnetic field changes with the K$_{\rm mex}$ index. K$_{\rm mex}$ is the analog of the Kp index but on a regional scale 
(three-hour measurements of the maximum absolute variations
of the horizontal component ($H$), \cite{pedro18}). 
In agreement with the Kp index, the values of K$_{\rm mex}$ run from 0-9, with "0" representing a quiet 
state and "9" representing 
a saturated value for the most disturbed states. Each K$_{\rm mex}$ level relates almost
logarithmically to 
$H$ deviations from its quiet baseline \citep{goodman2005}.

The quiet baseline is calculated by statistically removing the systematic diurnal and monthly variations of $H$. The algorithm of K$_{\rm mex}$ calculation obtains values consistent with those reported with its planetary counterpart (Kp). At present, for the K$_{\rm mex}$ calculation, we use data from the Teoloyucan magnetic observatory only; consequently, at this moment K$_{\rm mex}$ values are only regionally representative for central Mexico. In the future,  we will increase 
our coverage with several magnetometers in the Mexican territory. This will substantially increase the K$_{\rm mex}$ coverage as well our space weather monitoring capabilities. It is important to remark that K$_{\rm mex}$ is under development, and its data sets are nor definitive. Further details on  K$_{\rm mex}$ calculations can be found in \cite{pedro18} and general K-index calculations in \cite{1980GMS....22..607M} and \cite{1991RvGeo..29..415M}.




\subsection{GPS Receiver Stations}

There are different GPS receiver networks operating in Mexico \citep{2018ssn}. Data from these
networks are used to monitor SW effects on the ionosphere over 
the country. Vertical Total
Electron Content (TEC) values are obtained from RINEX files of local GPS stations with the
use of two methods.
In the first,
calculations are performed using the US-TEC software \citep{RDS:RDS5333,RDS:RDS5334}, wich
 is an
operational product at the Space Weather Prediction Center (SWPC), which is a product
developed through a collaboration between the National Geodetic Survey, SWPC by the National Oceanic
and Atmospheric Administration (NOAA), and the Cooperative Institute for Research in
Environmental Sciences of the University of Boulder, Colorado. US-TEC allows
performing near real-time monitoring of TEC over the region through the maps that  can be constructed in quasi real time.
For the detailed method description and its benefits, readers are referred to  \cite{RDS:RDS5333,RDS:RDS5334}.
 On the other hand, the TayAbsTEC method is also used for TEC calculations. Its advantage is that the receiver coordinates (specifics of the region) are taken into account.  
 Detailed description and benefits can be found in \cite{2015Ge&Ae..55..763Y} and \cite{SERGEEVA2017}. Both methods proved to provide satisfactory results for TEC estimation in the region; see references cited above.

The near real-time TEC maps over Mexico are one of the products developed by SCiESMEX. The
results of TEC estimation by different methods and the ionospheric W-index \citep{2013JASTP.102..329G} as a measure
of ionospheric disturbance are published in the SW weekly reports.




\subsection{Repository of Space Weather Data (RICE)}
In 2015, the National Council of Science and Technology 
created 
a network of repositories for science and technology in Mexico.
SCiESMEX manages the repository of the SW data. 
RICE provides the capabilities for massive and high-speed storage and processing of data from the networks of local and international SW instruments. The data in RICE can be processed in quasi real time. The results are published in the official SCiESMEX webpage both in weekly SW reports and as a quasi real-time SW values. This allows us to perform the continuous quasi real-time monitoring of SW conditions and to analyze previous events.   


\section{Space Weather Observations}

\subsection{The Event on 22 June, 2015}

On 19 June, 2015 at 05:00 UTC, a filament eruption was detected in the SSE quadrant of the solar disk. On 21 June, 2015, between 01:00 and 03:00 UTC, two solar flares erupted from the active region 2371 (M2 and M2.6), and a halo CME, associated with these flares,  was also detected. On the same day, at 09:44 UTC and 18:20~UTC, two other flares were released from the solar atmosphere, with M3 and M1 categories, respectively \citep{swpc21}. 

On 22 June, 2015, two CMEs arrivals were detected by ACE spacecraft at 04:51 UTC (associated with the first filament eruption) and 17:59 UTC (associated with the double peaked M2 flare from active region 2371 on 21 June). Shortly afterward, in the same active region, a M6 X-ray solar flare with a full halo CME was detected at 17:59 UTC. This last CME arrived to Earth two days later on 24 June at 12:58 UTC \citep{swpc22}.

\subsubsection{Observations by MEXART}
\begin{figure}
     \centerline{
\includegraphics[width=1.0\textwidth]{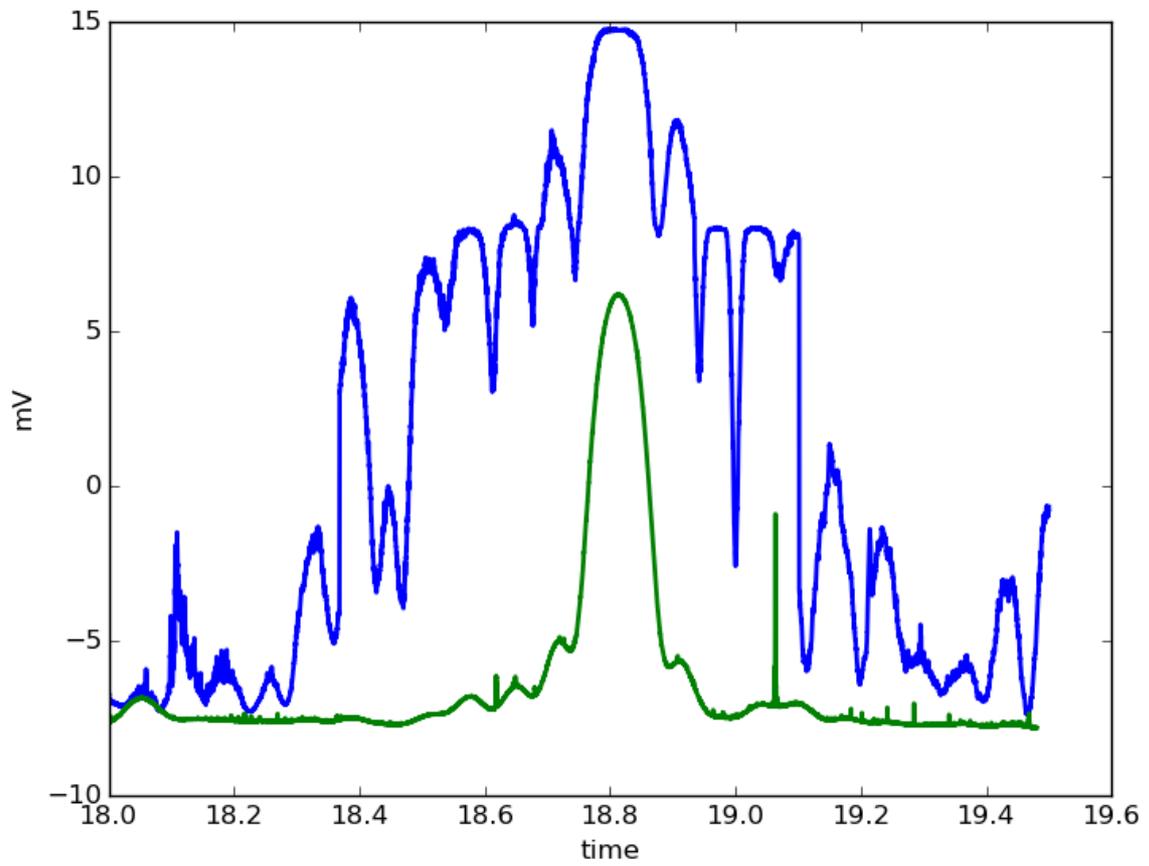}}
\caption{Two solar transits detected by MEXART. Electromagnetic flux measurements on June 22, 2015, during the occurrence of a solar flare (blue curve) and  regular quiet solar transit (green curve).}
\label{figure_quiet_sun_flare.eps}
\end{figure}

\begin{figure}
  \centerline{
\includegraphics[width=1.0\textwidth]{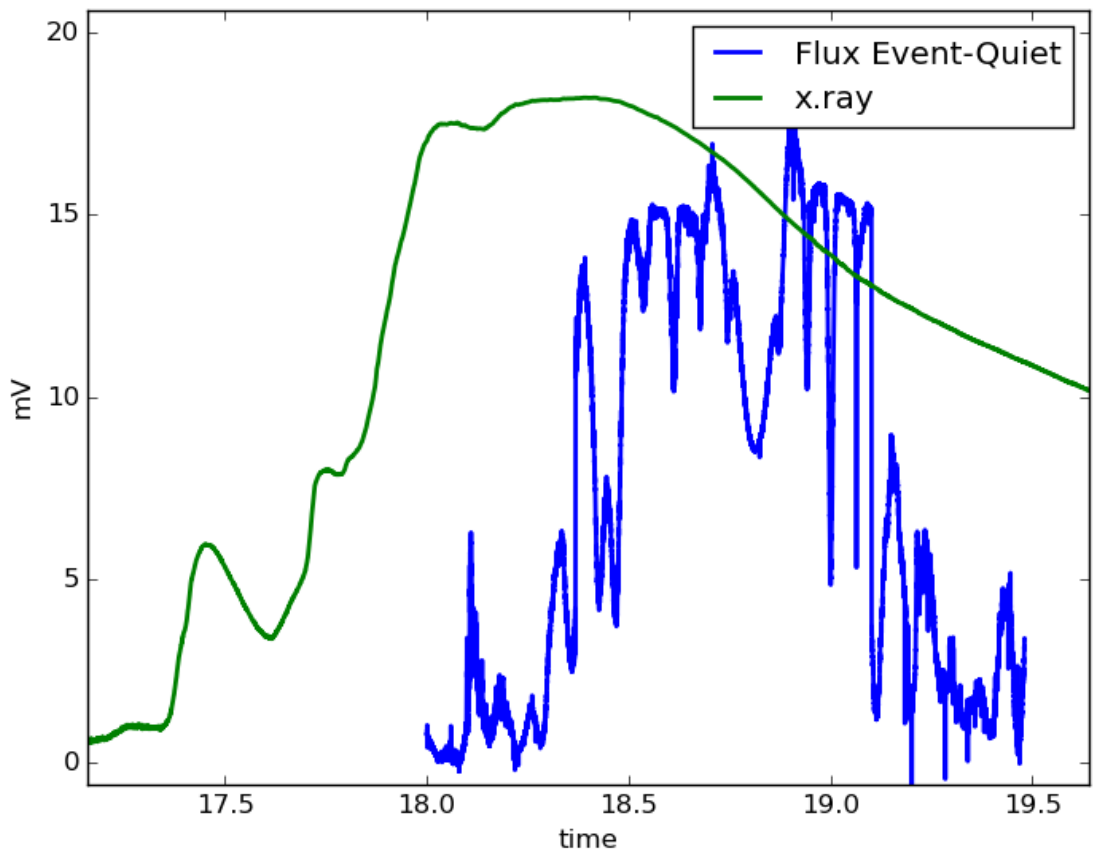}}
  \caption{Scaled X-ray flux from the solar flare June 22, 2015 by GOES satellite data (green curve) and a radio flux as detected by MEXART (blue curve).} 
\label{event-quiet+xray.eps}
\end{figure}

The Sun, as the strongest radio source in the sky, is detected daily by the MEXART.
These solar transit radio observations allow us to statistically characterize the flux and width of the Sun at 139.65 MHz and its variations 
within the solar cycle. 
It is possible to register solar activity or a flare during
the record of the solar transit, as ocurred on 22 June, 2015. 
During the M6.5 class flare, 
the Sun was 
near the local zenith around 22$^{\circ}$ in declination. 
Figure \ref{figure_quiet_sun_flare.eps} shows two records of solar transits 
obtained by MEXART: 
an example of a common quiet solar transit showing the expected radiation beam pattern of the antenna (green curve) and 
a solar transit registered immediately after the occurrence of the solar flare on 22 June, showing a disrupted and saturated beam pattern (blue curve). Figure~\ref{event-quiet+xray.eps} shows the 
signal of the X-ray flux increase of the solar flare detected by GOES (blue curve normalized to maximum MEXART flux) and the radio flux   
at 139.65 MHz detected by MEXART (green curve).  The temporal difference between the maximum of the X-ray emission detected by the GOES satellite and the solar transit detected by MEXART is about 30 minutes.
The flux in the MEXART data was saturated. 
Thus, it can be concluded that 
MEXART detected the signature of the solar flare
with a considerable increment in the radiation flux.

\subsubsection{Observations by CALLISTO}

During this event, the CALLISTO-MEXART station was still under initial configurations and did not detect the event.

\subsubsection{Geomagnetic Storm Detected with Local Magnetometer Data}
\begin{figure}
  \centerline{
\includegraphics[width=1.0\textwidth]{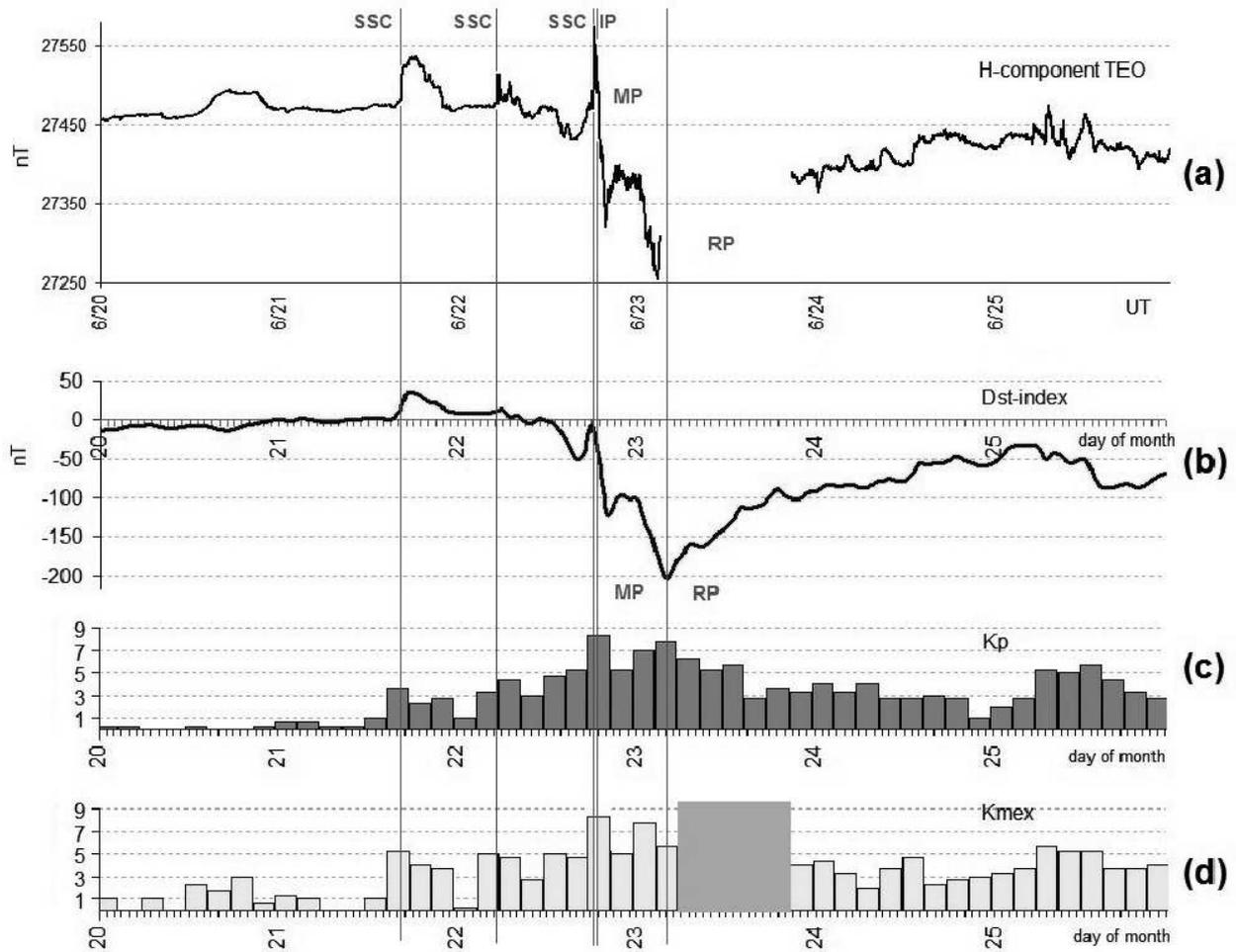}}
\caption{Variations of parameters during June 20-25, 2015: H-component of the magnetic field by
local magnetometer in Mexico (a), Dst-index (b), Kp-index (c) and local $K_{mex}$-index (d). Vertical
lines throughout all the panels indicate the moments of SSC, IP, MP and RP.}
\label{magnetic.eps}
\end{figure}

Figure \ref{magnetic.eps} shows the development of the geomagnetic storm from 20 to 25 June, 2015 (Figure \ref{magnetic.eps}a). Variations of the $H$-component of the magnetic field where measured at the Teoloyucan Geomagnetic Observatory (TEO). According to our magnetic data, three sudden storm commencements (SSC) provoked by the interplanetary shocks (Astafyeva et al., 2017) of different intensities occurred during this interval: 21 June at 16:46 UTC, 22 June at 05:47, and 22 June at 18:30~UTC.  An intense geomagnetic storm followed the last SSC with its main phase (MP) between 22 and 23 June. Unfortunately, some data where lost because the energy supply failed in the magnetic observatory between 3:33 UTC and 21:13 UTC on 23 June. For this reason, we show the Dst-index (Figure \ref{magnetic.eps}b) as the measure of geomagnetic field change on a global scale to define the beginning of the recovery (RP) of the geomagnetic field.

Figure \ref{magnetic.eps}c shows the Kp-index. The highest Kp values during the storm correspond to minimal $H$-component values. Figure \ref{magnetic.eps}d illustrates the values of K$_{\rm mex}$ derived from TEO data. K$_{\rm mex}$ reached its maximum value (K$_{\rm mex}=8+$) on 23 June, 2015 during the main phase of the storm. The gray area in the plot means TEO data was missing. Consider that currently the local geomagnetic K$_{\rm mex}$ index is undergoing a validation process. The results shown in Figure \ref{magnetic.eps}c are considered preliminary. In the near future, we will increase the number of magnetometers at different sites in the country. This will produce a better coverage to comparing the geomagnetic response in different regions in Mexico.



\subsubsection{Cosmic Rays Observations}

\begin{figure}
  \centerline{
\includegraphics[width=1.0\textwidth]{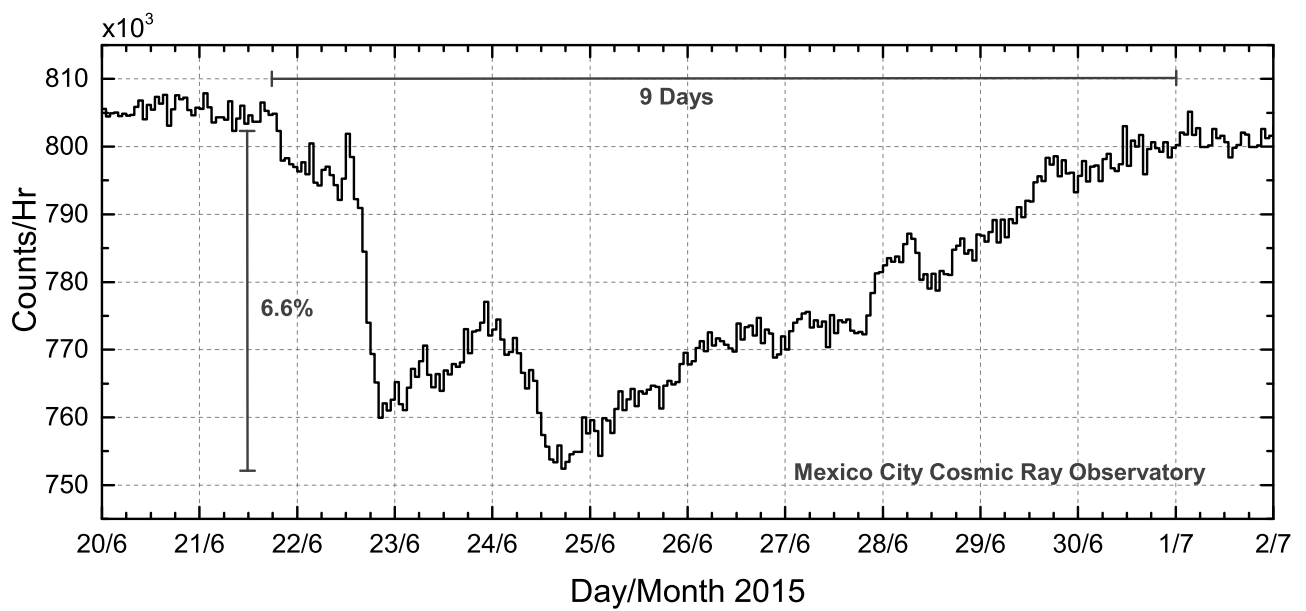}}
\caption{Forbush decrease detected by the Mexico City Cosmic Ray Observatory, generated by the M-class solar flares on June 21 to 22.}
\label{Forbush_Decrease_22-06-2015.eps}
\end{figure}

Figure \ref{Forbush_Decrease_22-06-2015.eps} shows the FD observed by the Mexico City NM. From the baseline, the minimum of the FD is estimated as 6.6\%. The interval to reach the minimum was about 74 hours, and the whole event lasted about 10 days. The event began on 22 June at 01:00~UTC. There was an abrupt flux decay on 22 June at 19:00~UTC, and it reached the first minimum on 23 June at  03:00~UTC, covering a decrease of at least 5\% in an interval of about eight hours. This main phase of the FD might well be associated with the arrival of the CME on 22 June at 17:59 UTC.  The behavior of the cosmic rays fluxes correlated with the geomagnetic disturbances described in Figure \ref{magnetic.eps}.
Other NM stations around the world also confirmed this FD event. This event shows that the
Mexico City NM works properly, and it detects the variations of cosmic rays associated with SW events.
Unfortunately, the muon telescope at Mexico City was off at the time of the event.


\subsubsection{Ionospheric Disturbances Detected with Use of GPS Data}

\begin{figure}
 \centerline{
\includegraphics[width=1.0\textwidth]{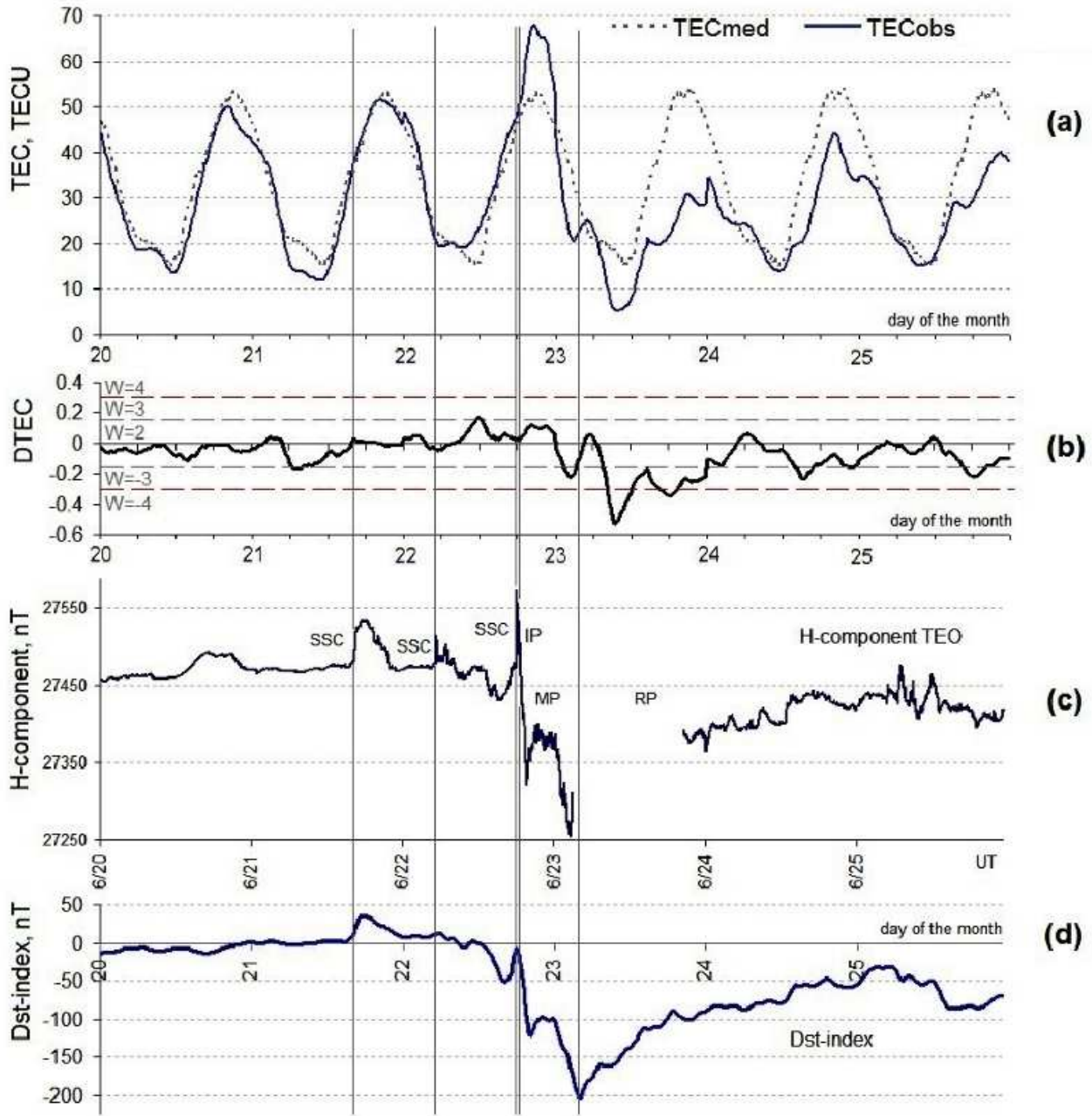}}
\caption{Ionospheric and geomagnetic parameters during June 20-25, 2015: observed (TECobs)
and 27-day median (TECmed) TEC values for UCOE station (a); ionospheric weather W-index
corresponding to the logarithmic deviation DTEC for UCOE station (b); H-component (c) and Dst-
index (d) variations.}
\label{kpmex.eps}
\end{figure}

The discussed geomagnetic storm between 22 and 23 June, 2015, caused ionospheric disturbances over Mexico.
First, let us consider the data from a single GPS receiver. Figure \ref{kpmex.eps}a
illustrates the behavior of one of the main ionospheric parameters, TEC, during 20 and 25 June, 2015. The values of the observed TEC (blue curve) and it 27-day running median (gray dotted curve) are shown for the GPS station UCOE located at the MEXART site (latitude 19.8 N, longitude 101.68 W). Median TEC values serve as a quiet reference. The TEC values observed from 20 to 21 June, 2015, followed a quiet pattern (Figure \ref{kpmex.eps}a).
The difference between the observed and median TEC curves during these days is within the day-to-day variability limits \citep{SERGEEVA2017}. In contrast, during the geomagnetic storm, TEC showed the positive phase of the ionospheric disturbance, which commenced exactly with the beginning of the main phase of the storm at 18:30 UTC on 22 June (see vertical lines throughout the panels of Figure \ref{kpmex.eps}). TEC then showed the negative phase of disturbance with the beginning of the recovery phase of the storm approximately at 04 UTC on 23 June and further during the next days. The full TEC recovery, to its quiet level, did not occur until 28 June, 2015 (not shown for the economy of space). These results are in line with other ionospheric studies of this event \citep{2017JGRA..12211716A}. We used the ionosphere weather W-index \citep{2013JASTP.102..329G,blago2018}
to estimate the intensity of the ionospheric disturbance. W-index specifies the ionosphere from a quiet state to an intense storm. The categories of W-index correspond to the logarithmic deviation of TEC from its quiet median:
$$DTEC = log (TECobs/TECmed),$$
where $TECobs$ is the observed TEC value and $TECmed$ is a median value calculated over 27 days prior to the day of observation (blue and gray dotted curves in Figure \ref{kpmex.eps}a respectively). The value $DTEC$ < $-0.301$ corresponds to $W = -4$. According to the classification provided in \cite{2013JASTP.102..329G}, this is the intensely negative W-storm. $W = -4$ represents the least negative storm magnitude. Such a disturbed state of the ionosphere can lead to negative impact on different systems. For example, the range of the used frequencies within the HF band can narrow significantly. Such negative storms are especially dangerous during the night hours when critical frequencies of the ionosphere are lower than during the daytime. This was our case: the most intense negative $DTEC$ minimum on 23 June (Figure \ref{kpmex.eps}b) occurred during the local night hours.

\begin{figure}
 \centerline{
\includegraphics[width=1.0\textwidth]{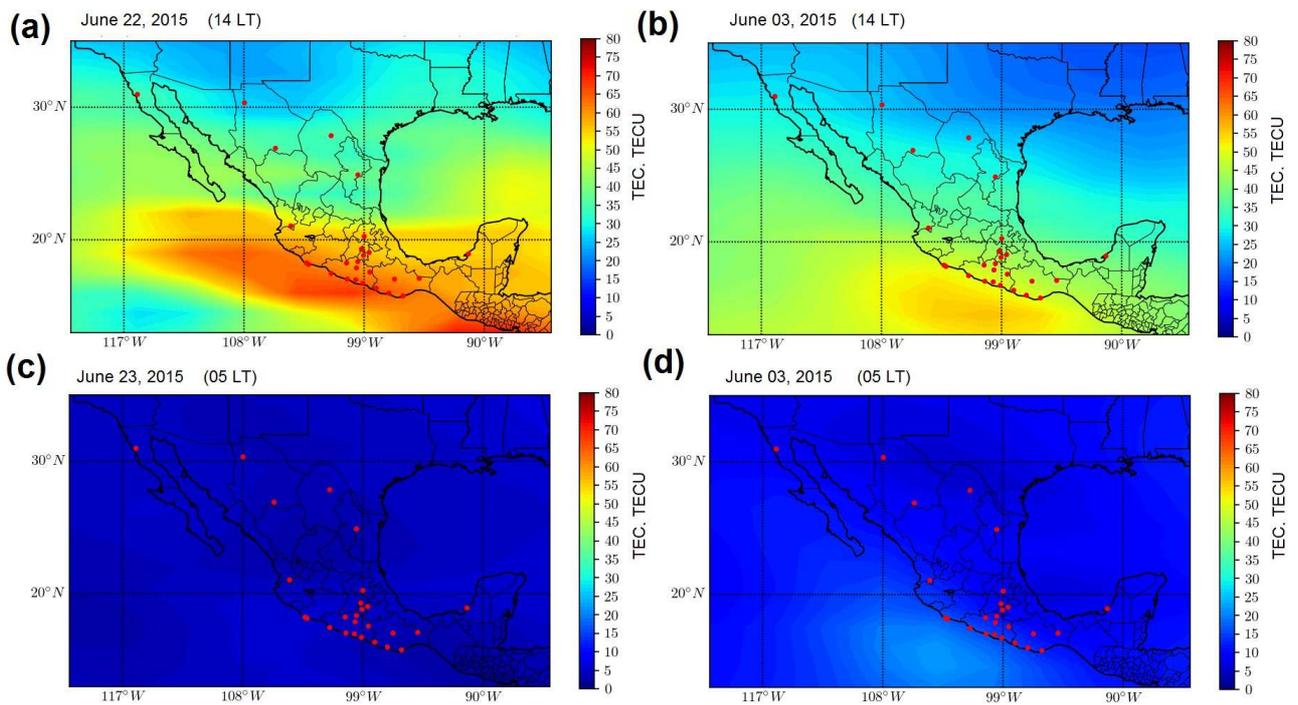}}
\caption{Regional TEC maps constructed for two time moments during the disturbance (panels a and c) and for the same moments under quiet geomagnetic conditions (panels b and d). The location of GPS stations whose data was used are marked by red points (for more details, we have prepared a video in a period of 38 days from 2015/05/24 to 2015/06/30 at 24 FPS at \url{http://www.rice.unam.mx/aztec/videos/tecmaps01.mp4}).}
\label{msergeeva1.eps}
\end{figure}

One of the products that SCiESMEX offers to its users is the regional TEC maps, which illustrate TEC distribution over Mexico. Such maps are a useful instrument for qualitative estimation of the ionosphere state, for instance \citep{2009JGeod..83..263H,2017AdSpR..60.1606R,2018AnGeo..36...91B,articleli}. The example is provided in Figure \ref{msergeeva1.eps}. Data from 22 GPS receivers were used to construct these TEC maps. The main diurnal TEC maximum is usually observed near 14 local time (20 UTC) throughout Mexico during all four seasons \citep{SERGEEVA2017,SERGEEVA2018}.

The results for 14 local time on the day of the maximum positive TEC disturbance on 22 June, 2015 are compared to the results for the same hour on the
quiet geomagnetic day of 3 June, 2015 (Figure \ref{msergeeva1.eps}a,b). The map for 3 June, 2015 can serve as a quiet reference as it was
one of the quietest days of the month (Dst min=-4 nT, Kp max=2-, $K_{\rm mex}$ max=3). The moment of 14~LT was also a moment of
the largest difference (positive disturbance) between the disturbed and quiet TEC values during the considered ionospheric
disturbance\footnote{For more details we have prepared a video of a 38-day period from 24 May, 2015 to 30 June, 2015 at 24 FPS at \url{http://www.rice.unam.mx:8080/aztec/videos/tecmaps01.mp4}}. If one compares the plots in Figure \ref{msergeeva1.eps}a and Figure \ref{msergeeva1.eps}b, we see that the picture is rather different: the electron concentration
in the ionosphere was increased over all Mexico on the day of the storm. TEC growth can lead to the increase
of errors in the object location with signals of global positioning systems. The difference between these two maps proves that
the ionosphere structure changed significantly during the disturbance. In addition, we illustrate TEC distribution maps
constructed for 05 LT (11 UTC) on 23 June, 2015 (the moment of the most intensely negative TEC disturbance), and for the same
moment of the quiet day on 3 June, 2015 (Figure \ref{msergeeva1.eps}  panels c, d). Clearly, the TEC was lower on 
this
day.

As TEC is an integral parameter that characterizes electron content in the cross unit section from the ground to a GPS satellite, it does not permit making precise conclusions about the variations in different ionospheric layers (their peak density and height). The ionospheric sounding data are usually used for that. There are no ionosonde measurements in Mexico at the moment. Installation is a part of our future work. To sum up, the geomagnetic storm that started on 22 June, 2015, provoked the ionospheric disturbance over Mexico, which was characterized by positive and then by negative phases. These phases of ionospheric disturbance correlated with the phases of geomagnetic disturbance. The structure of the ionosphere was significantly changed during the geomagnetic storm, which could lead to negative consequences for different technological systems.

\subsection{The Event on 26 September, 2015}\label{mexcal}
The second event that we address in this study is related to the solar radio burst.
It was detected by the CALLISTO-MEXART station and the MEXART radio telescope at 19:22 UTC on 2 September, 2015.
This solar radio burst was associated with a weak M1.1 solar flare that started at 19:20 UTC, peaked at 19:24 and ended at 19:27 UTC.
The closest CME related to this event was recorded by LASCO at 20:00:04 UTC. According to SWPC/NOAA this particular CME did not hit the Earth \citep{swpc2092}.

\begin{figure}
 \centerline{
\includegraphics[width=1.0\textwidth]{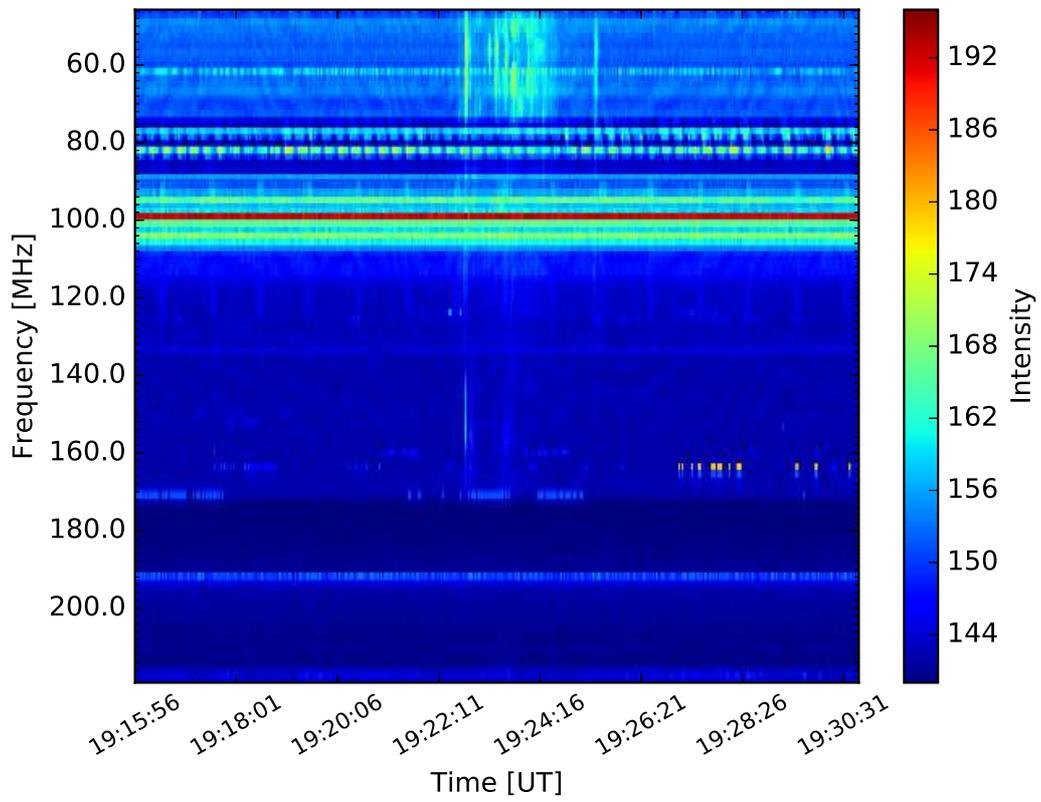}}
\caption{Type III solar radio burst detected by CALLISTO-MEXART on September 29, 2015.}
\label{CALLISTO-20150929.eps}
\end{figure}
\begin{figure}
 \centerline{
\includegraphics[width=0.9\textwidth]{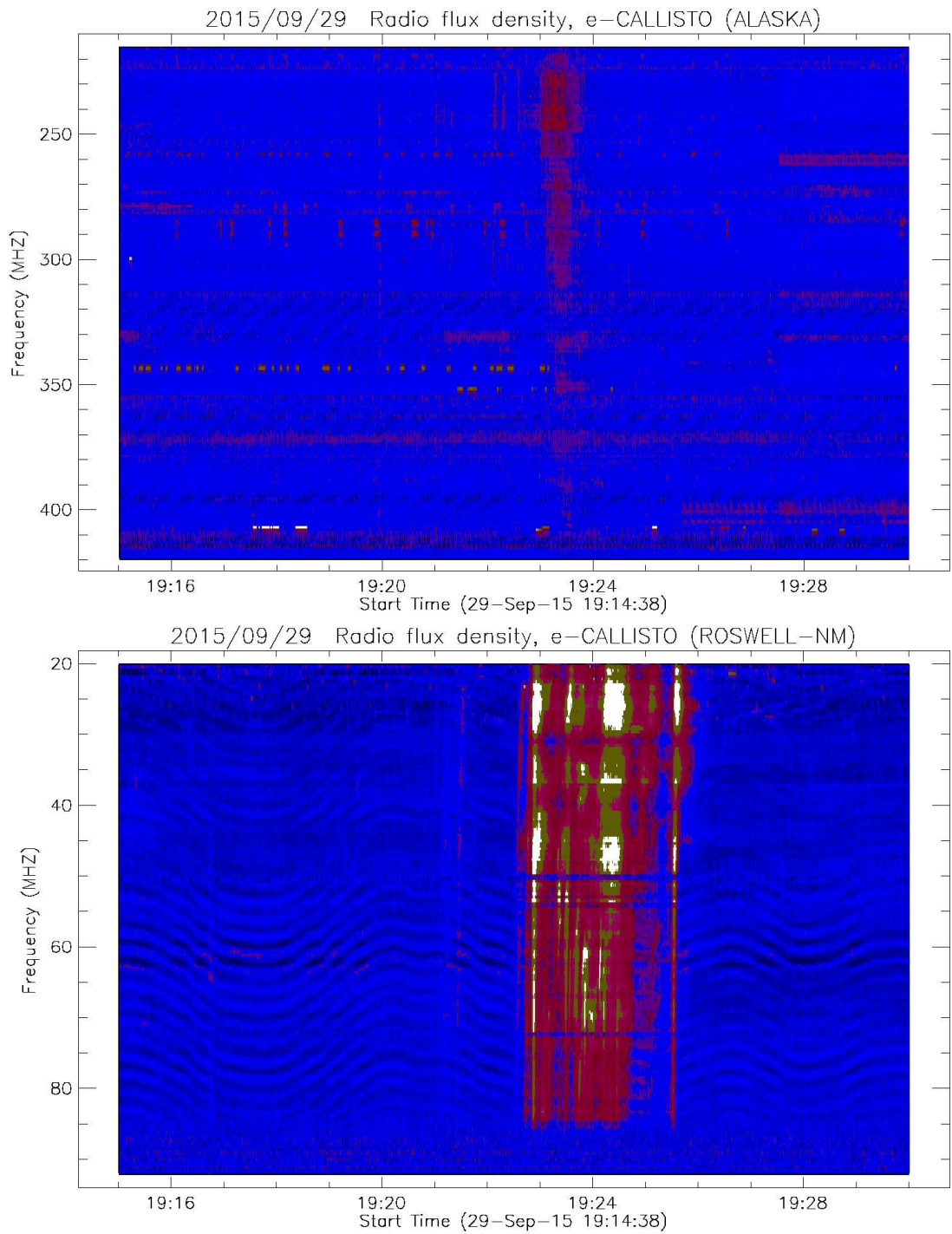}}
\caption{Type III solar radio burst registered by e-callisto station on September 29, 2015. Upper panel: Alaska Station; Buttom panel: ROSWELL-NM station. In both records the radio burst is registered at 19:23 UTC.}
\label{bothpanelscallisto.eps}
\end{figure}
\begin{figure}
 \centerline{
\includegraphics[width=1.0\textwidth]{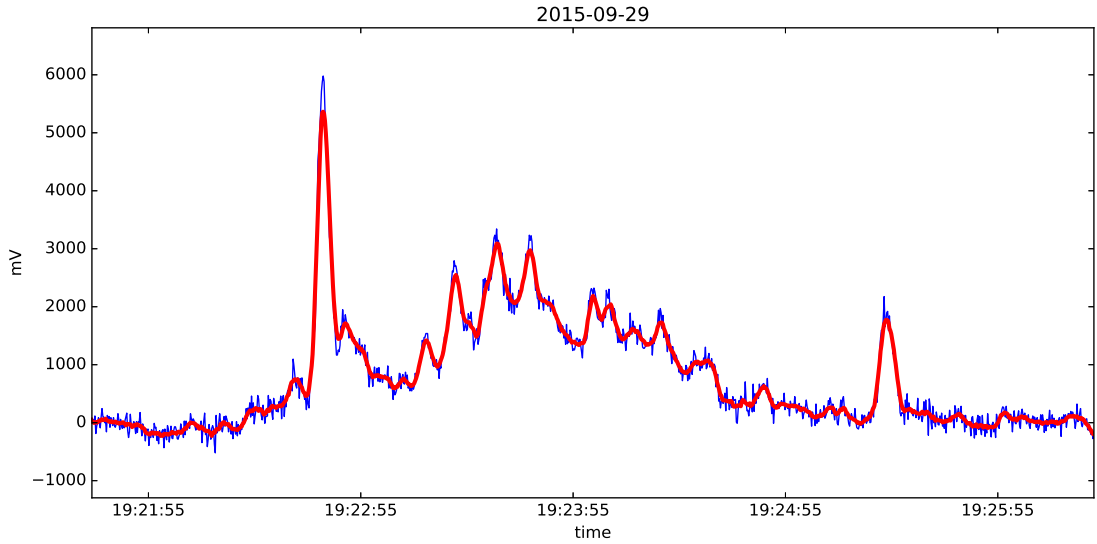}}
\caption{Light curve of the type III solar radio burst detected September 29, 2015 with signal-to-noise ratio $= 32$.}
\label{lightcure-20150929.eps}
\end{figure}

The solar radio event registered by CALLISTO-MEXART (Figure \ref{CALLISTO-20150929.eps}) shows the characteristic dynamic solar spectrum of a radio burst type III. The type II and type III spectrums differ in their dynamics. While type III shows an emission in consecutive short periods (minutes) of time and covers a wide wavelength bands, the type II radio burst derives from high to low frequencies for a long period of time (days) \citep{JGRA:JGRA21437}. Related to this event, SWPC releases an ALTTP2 code where a type II at 19:30 UTC was reported (serial number: 1025). The type III solar radio burst was confirmed by two independent CALLISTO stations: Roswell-NM and Alaska (Figure \ref{bothpanelscallisto.eps}).
Figure \ref{lightcure-20150929.eps} shows the light curve of the event
constructed on the basis of CALLISTO data.
The maximum peak of the event 
was characterized by the signal-to-noise ratio (SNR) of 32.

Both instruments MEXART radio telescope and the CALLISTO-MEXART have a common band: approximately 140 MHz. 
Consequently, both instruments
can 
detect the same events.

\begin{figure}
 \centerline{
\includegraphics[width=1.0\textwidth]{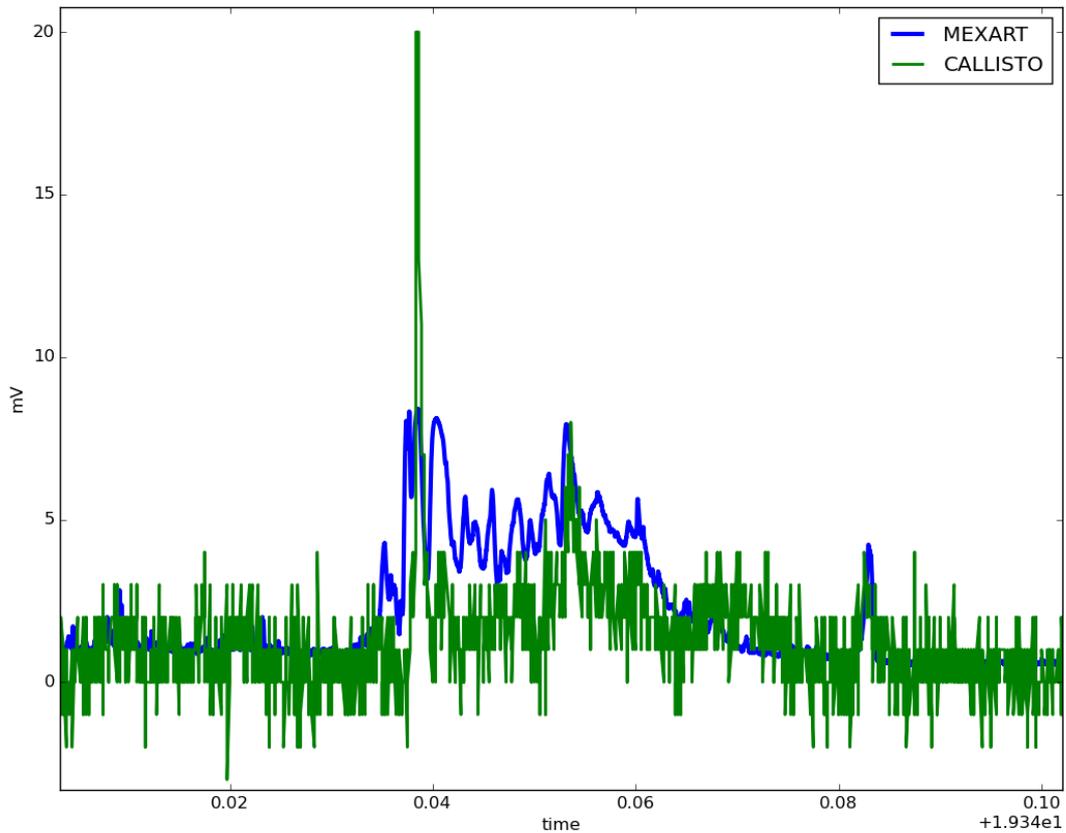}}
  \caption{September 29, 2015 records by MEXART (blue curve) and CALLISTO-MEXART (green curve) at approximately 140 MHz. The flux was scaled in order to compare the response of both instruments. 
  }
\label{MEXART_CALLISTO_event.eps}
\end{figure}

We rescaled (y-axis) the fluxes obtained by two instruments to compare them. The results are given in Figure~\ref{MEXART_CALLISTO_event.eps}
where we show the event observed at the same band in both instruments.
We see that the solar radio burst was detected by both instruments.
This is the first record of a solar radio burst type III
by the MEXART radio telescope
confirmed by the CALLISTO station simultaneous observation.
The flux registered by both instruments opens the possibility of using MEXART as a solar radio monitor (not only for IPS observations) with high time resolution and a better sensitivity. The SNR only for the 140 MHz channel of CALLISTO is 18, and for the MEXART, it is 224 for the closest band). 

The records shows several disturbances in radio communications between 50 and 75 MHz and radio noise between 110 and 170 MHz for around two minutes at the site of observation. This is the first spectrum that shows a radio blackout over Mexico related to SW.
This radio blackout could probably affect the frequencies lower than 50 MHz. The local time of the radio burst was about 11:20, close to noon locally. The ionosonde measurements could provide us with information if the HF band (3-30MHz) was affected by the event. As mentioned above, SCiESMEX had no ionosonde measurements in Mexico at that moment. The installation of the ionosondes for oblique ionospheric sounding is planned as our future work. 

\section{Conclusions}
We presented the results of the first joint observations of SW phenomena in Mexico. We addressed two SW events that occurred on 22 June, 2015 and 29 September, 2015. Features of the behavior of SW parameters were obtained with the use of different local instruments installed in Mexico. The main results are the following:

\begin{itemize}
\item
A solar flare was detected by the MEXART radio telescope on 22 June, 2015, in agreement with GOES satellite data. This example proves the possibility of using MEXART for solar flare detection if they occur during the local daylight hours.

\item
We presented for the first time a solar radio event (29 September, 2015) detected by the MEXART radio telescope and confirmed by the CALLISTO-MEXART station. The measuments by CALLISTO-MEXART were in accord with other CALLISTO observations. This proves that both ground-based local instruments (MEXART and CALLISTO-MEXART) can be used for the monitoring of solar radio bursts, which occur during the local daylight hours in Mexico. The advantage of the MEXART instrument is better sensitivity for such events. Note also that we report, for the first time, a radio blackout over Mexico related to SW phenomena.

\item
Local cosmic rays data indicate SW phenomena in Mexico. This is due to the fact that the irregularities in the interplanetary magnetic field, associated with large-scale solar wind disturbances, deflect the cosmic ray flux measured in the center of Mexico. For example, the record of Forbush decrease associated with the passing of the CME detected during the event in June 2015.

\item
Local geomagnetic field variations from 21 to 25 June, 2015 caused an intense ionospheric disturbance over Mexico. Local magnetometer data were in accord with the variations of the Dst index. The regional K$_{\rm mex}$ index allowed us to estimate the rate of geomagnetic disturbance in Mexico. The phases of ionospheric disturbance correlated with the phases of geomagnetic disturbance in time. The results are in agreement with other ionospheric studies of this event. It was revealed that the structure of the ionosphere was significantly changed during the geomagnetic storm which could lead to negative consequences for different technological systems.
As the ionosphere state was estimated with only TEC data, no conclusions about the changes in each ionospheric layer can be made. 

\end{itemize}

Some lessons can be learned from this first study in order to enhace the SW monitoring and the development of a comprehensive ground-based multi-instrument data set in Mexico. We must increase the number of magnetometers, located at different sites, to have local measurements at different regions in real time. The installation of more CALLISTO stations in Mexico will allow us to understand the effects of radio comunications disruption with more accuracy. For the case of TEC maps computed over Mexican territory, the next step is to improve the spatial resolution of TEC maps by increasing the number of GPS stations available and by bettering our TEC calibration methods. One of the future steps for improving the computations of TEC maps is to implement a homogeneous distribution of GPS stations throughout the ground territory.
More detailed analysis could be made with the ionospheric sounding data by ionosondes. Thus, the ionosonde data are needed to complement both radio blackout studies and ionospheric radio propagation conditions over Mexico.  The incorporation of a network of magnetometers and ionosondes in Mexico in the next year will significantly improve the coverage and quality of our space weather data.

\section*{acknowledgements}
Thanks are expressed to Catedras CONACyT
(CONACyT Fellow) for supporting this work. Victor De la Luz
acknowledges CONACyT 254497 and CONACyT 268273 for
Ciencia Basica and Repositorios Institucionales. Maria A. Sergeeva
acknowledges the funding by CONACyT-AEM 2017-01-292700.
Julio C. Mejia-Ambriz acknowledges CONACyT 256033.
Pedro Corona-Romero acknowledges CONACyT 254812.
SCiESMEX is partially funded by CONACyT-AEM grant 2017-
01-292684, CONACyT LN 293598, CONACyT PN 2015-173,
and DGAPA-PAPIIT IN106916. Ernesto Aguilar-Rodriguez
acknowledges the DGAPA-PAPIIT project (grant: IN101718) and
the CONACyT project (grant: 220981). Mario Rodriguez-Martinez
acknowledges DGAPA-PAPIIT IA 107116 and CONACyT
INFR: 253691. The authors express their gratitude to the NOAA
Space Weather Prediction Center (SWPC), Boulder, Colorado,
USA, for providing the analysis software used at SWPC for the
operational US-TEC product to perform TEC calculations for this
study. The calculations of the local TEC values are partly based on
GPS data provided by the Mexican Servicio Sismológico Nacional
(SSN, 2018; Pérez-Campos et al., 2018), the Trans-boundary,
Land and Atmosphere Longterm Observational and Collaborative
Network (TLALOCNet; Cabral-Cano et al., 2018), and
SSN-TLALOCNet operated by the Servicio de Geodesia Satelital
(SGS) and SSN at the Instituto de Geofísica, Universidad Nacional
Autónoma de México (UNAM) and UNAVCO Inc. We gratefully
acknowledge all the personnel from SSN, SGS, and UNAVCO
Inc. for station maintenance, data acquisition, IT support, and data
distribution for these networks. TLALACNet, SSN-TLALOCNet,
and related SGS operations are supported by the National Science
Foundation, grant number EAR-1338091, NASA-ROSES
NNX12AQ08G, Consejo Nacional de Ciencia y Tecnologia
(CONACyT) projects 253760, 256012, and 2017-01-5955, UNAM
Programa de Apoyo a Proyectos de Investigación e Innovación
Tecnológica (PAPIIT) projects IN104213, IN111509, IN109315-3,
IN104813-3, and supplemental support from UNAM Instituto
de Geofísica and Centro de Ciencas de la Atmosfera. Thanks
are expressed to the Institute for Astronomy, ETH Zurich, and
FHNW Windisch, Switzerland. Whitham Reeve and Stan Nelson
are thanked for providing the observations for the stations of
Alaska and Roswell, New Mexico, in the e-CALLISTO network.
The authors would like to thank Ana Caccavari for providing the
magnetic field data from Teoloyucan Geomagnetic Observatory.
The authors also thank Ilya Zhivetiev from the Institute of Cosmophysical
Research and Radio Wave Propagation FEB RAS
and Yury Yasyukevich and Anna Mylnikova from the Institute of
Solar-Terrestrial Physics SB RAS for providing the TayAbsTEC
software for this study (http://www.gnss-lab.org/tay-abs-tec; last
access: 30 June 2018).

\newpage
\newpage
\newpage
\newpage
\newpage
\newpage
\clearpage

\end{document}